\documentclass[12pt]{article}

\usepackage{scicite}
\usepackage{graphicx}
\usepackage{xcolor}
\usepackage{soul}

\usepackage{times}

\newenvironment{sciabstract}{%
\begin{quote} \bf}
{\end{quote}}

\title{Density, not radius, separates rocky and water-rich small planets orbiting M dwarf stars}

\author
{Rafael Luque$^{1,2,\ast}$ \& Enric Pallé$^{3,4,\dagger}$\\
\\
\normalsize{$^{1}$Instituto de Astrof\'isica de Andaluc\'ia }\\
\normalsize{(Consejo Superior de Investigaciones Cient\'ificas), Granada 18008, Spain.}\\
\normalsize{$^{2}$Department of Astronomy \& Astrophysics, University of Chicago, Chicago, IL 60637, USA.}\\
\normalsize{$^{3}$Instituto de Astrof\'isica de Canarias, La Laguna, Tenerife 38205, Spain.}\\
\normalsize{$^{4}$Departamento de Astrof\'isica, Universidad de La Laguna, La Laguna, Tenerife 38206, Spain.}\\
\\
\normalsize{$^\ast$Corresponding author. E-mail: rluque@uchicago.edu.}
\\
}

\date{}

\begin{document} 


\baselineskip24pt


\maketitle


\begin{sciabstract}


Exoplanets smaller than Neptune are common around red dwarf stars (M dwarfs), with those that transit their host star constituting the bulk of known temperate worlds amenable for atmospheric characterization. We analyze the masses and radii of all known small transiting planets around M dwarfs, identifying three populations: rocky, water-rich, and gas-rich. Our results are inconsistent with the previously known bimodal radius distribution arising from atmospheric loss of a hydrogen/helium envelope. Instead, we propose that a density gap separates rocky from water-rich exoplanets. Formation models that include orbital migration can explain the observations: Rocky planets form within the snow line, whereas water-rich worlds form outside it and later migrate inward.

\end{sciabstract}

\newpage

Exoplanets which transit red dwarf stars (also known as M dwarfs) intercept a large fraction of the stellar disk, making them potentially suitable targets for transmission spectroscopy studies \cite{Rauer2011}. The habitable zones of planetary systems around M dwarfs are located close to the host star, increasing the chance of transits occurring. Whether small planets around M dwarfs are potentially habitable remains unclear, in part due to incomplete knowledge of their composition \cite{Segura2005}. 

Small exoplanets are known to have a bimodal radius distribution, with two populations separated by a gap known as the radius valley \cite{Fulton17}. Potential explanations focus on atmospheric mass loss mechanisms, such as photoevaporation triggered by the host \cite{OwenWu2017ApJ...847...29O,JinMordasini2018ApJ...853..163J} or from the internal heating of the planet \cite{Ginzburg2018MNRAS.476..759G}. Photoevaporation models can reproduce the position of the radius valley by assuming that super-Earth and sub-Neptune planets all have a rocky composition, with their different radii being a consequence of whether or not they retain their primordial hydrogen/helium atmosphere (H/He envelope). If the internal composition of these planets was icy, the radius valley would be at larger planetary radii \cite{OwenWu2017ApJ...847...29O,RogersOwen2021MNRAS.503.1526R}.

A purely rocky composition for most short-period small exoplanets is inconsistent with global formation models that include accretion and migration mechanisms \cite{Raymond2018}. These predict that planets with masses below 20\,Earth\,masses ($M_\oplus$) become water rich, because large planet embryos are preferentially formed beyond the ice line (at a distance from the central protostar where volatiles are cold enough to condense into solid ice grains) and migration dynamics efficiently moves objects in this mass range inwards \cite{Mordasini2009a}. These models reproduce other observed features of the small exoplanet population such as the period ratio distribution of adjacent planet pairs and the over-abundance of single transiting systems \cite{Izidoro2017MNRAS.470.1750I}. Those results were based upon observations of planet radii alone. The density of each planet might provide more information, but it requires measurements of both mass and radius.

We investigate the population of small  (radius of the planet smaller than 4 Earth radii, $R < 4\,R_\oplus$) transiting planets around M dwarfs (hereafter STPMs). Determining the masses of planets observed by transiting missions requires ground-based follow-up, with some planets having multiple estimates in the literature. We compiled all published mass measurements for STPMs and used archival observations to refine the physical parameters of nine planets in seven planetary systems (Table~S8). We use the \textsc{juliet} code \cite{juliet} to model transits and radial velocities \cite{Supplementary}. To build our sample of STPMs, we began with the Transiting M-dwarf Planets catalogue\cite{Trifonov2021Sci...371.1038T}, which includes 43 planets with a radius smaller than $4\,R_\oplus$ in 26 planetary systems as of July 21, 2021. We restrict our analysis to those which are precisely characterised, which we consider as dynamical mass precision better than 25\% and radius precision better than 8\%. After our analysis of archival observations \cite{Supplementary}, 34 of the planets (80\%) are precisely characterised by this definition.

Figure~1A shows our STPM sample on a mass-radius diagram, compared with theoretical composition models\cite{Zeng2019PNAS..116.9723Z}. We find that the planets do not form a continuum, but are distributed in three separate populations. Two groups are consistent with specific compositions: the extrapolated mass-radius relation of Earth (hereafter rocky planets); and planets consisting of rock and water ice in 1:1 proportion by mass (hereafter water worlds). The third group consists of planets with larger radii than either model, requiring H/He envelopes. We assign each planet to the closest model composition, taking into account the uncertainties in mass and radius. In Fig.~1B, we show the same sample in a mass-density diagram, where the bulk densities of the planets have been normalised by a theoretical model of an Earth-like composition (scaled Earth's bulk density, $\rho_{\oplus,s}$: 32.5\% iron mass fraction, 67.5\% silicates) which accounts for gravitational compression \cite{Zeng2019PNAS..116.9723Z}.

The rocky population spans a large range of equilibrium surface temperatures ($T_{eq}$) and has a small dispersion in density. Many of these planets are close enough to their host stars to experience strong runaway greenhouse effects, so they are candidates to have extended atmospheres of water in supercritical state \cite{Turbet2020, Mousis2020}. With little liquid water on the surface, this increases the planetary radius compared to water-free planets (Fig.~2D). These planets must have thin or non-existent H/He atmospheres or supercritical water layers. Otherwise, small variations in the mass fraction of H/He envelopes (Fig.~2A) or in their temperatures (Fig.~2C)  would result in large differences in the radius of the planets\cite{Zeng2019PNAS..116.9723Z,Turbet2020,Mousis2020}. Therefore, these planets aligned must be water-rich objects, not gas-rich.

The third population have radii larger than $2.3\,R_\oplus$ and masses higher than $6\,M_\oplus$. These are larger than rocky or water-rich planets of the same mass, so we refer to them as puffy sub-Neptunes. The nature of this population is more difficult to determine, because interior and atmospheric composition models are degenerate. The possible scenarios include rocky worlds with massive H/He envelopes (Fig.~2A) or water worlds with thin envelopes (Fig.~2B), perhaps affected by a greenhouse effect that generates extended atmospheres of water in super-critical state\cite{Mousis2020}. However, there are no differences in $T_{\rm eq}$ between the water world and puffy sub-Neptunes populations, and nearly all planets have $T_{\rm eq} > 400\,\mathrm{K}$, high enough to potentially have inflated hydrospheres. Therefore, the larger radii dispersion of puffy planets could be a consequence of the individual H/He accretion histories, not atmospheric loss processes. If so, the water worlds and puffy sub-Neptunes could be part of a continuous population, with differences in their bulk densities arising from their different masses, which affect their accretion potential. Observations of water in the transmission spectra of the puffy sub-Neptunes K2-18~b\cite{Tsiaras2019NatAs...3.1086T,Benneke2019ApJ...887L..14B} and HD~106315~c \cite{Kreidberg2020} are consistent with this scenario, as is the presence of a thick atmosphere that likely contains water in $\pi$~Mensae~c\cite{GarciaMunoz2021}.

Growth models and Monte Carlo simulations have led to similar conclusions for water worlds\cite{Zeng2019PNAS..116.9723Z}. An issue with this interpretation is that water worlds with masses between 3--6\,$M_\oplus$ should have radii 1.5--2.0\,$R_\oplus$, so no radius valley would be observed \cite{Venturini2020}. Figure~1 shows that rocky planets exist from 0.3 to about 10 Earth masses. The radius valley for M dwarfs is not empty: the apparent scarcity of small planets with radii between 1.5--2.0\,$R_\oplus$ is due to a combination of the rocky population having a maximum mass of $10\,M_\oplus$ and the water worlds a minimum mass of 2--3\,$M_\oplus$, and the corresponding radius limits for each (see Supplementary Text).

Figure~3 shows normalised radius and density histograms for the STPM sample, which we fitted with Gaussian functions.  We find mean bulk density (radius) in each population of $0.94\pm0.13\,\rho_{\oplus,s}$ ($1.21\pm0.28\,R_\oplus$) for rocky planets, $0.47\pm0.05\,\rho_{\oplus,s}$ ($1.97\pm0.28\,R_\oplus$) for water worlds, and $0.24\pm0.04\,\rho_{\oplus,s}$ ($2.85\pm0.63\,R_\oplus$) for puffy sub-Neptunes. For rocky planets the average radius is limited by the detection limit of transiting surveys. The separation between super-Earths and sub-Neptunes, which was roughly determined at $1.6\,R_\oplus$ from the radius valley \cite{CloutierMenou2020AJ....159..211C}, is more pronounced in density: we find a clear separation at $0.65\,\rho_{\oplus,s}$ with no overlap between populations.

Figure~4 shows the radius-period and density-period diagrams. We used the code \textsc{gapfit}\cite{gapfit} to determine the location and slope of the gap that separates rocky planets from water worlds in these two representations. For our STPM sample, we find the orbital period has no dependence on radius (slope 0.02$\pm$0.04). In addition, we find no significant dependence (all slopes are consistent with zero within $1\sigma$) of planet density on orbital period, incident bolometric flux $S$ or stellar mass $M_\star$ (Figs.~4~\&~S18).

Our identification of three classes of STPMs according to their bulk density is consistent with formation and evolution theories. Accretion mechanisms predict that ice and rock both participate in planetary growth. Material condensing beyond the water ice line is expected to have a 1:1 water-to-rock ratio if it has the same composition as the Solar System\cite{Lodders2003}, the same ratio we used in Fig.~1. However, the STPM sample has no planets with intermediate water-rock compositions, which are predicted in planetesimal accretion models but not in pebble accretion models \cite{Brugger2020,Venturini2020}. Therefore, our classification favours pebble accretion models as the main mechanism for forming small planets around M dwarfs. Synthetic population of small planets around M dwarfs\cite{Burn2021arXiv210504596B} predict a mass-radius diagram which is consistent with our STPM sample (Fig.~S20). The simulations predict that water worlds are more common at lower stellar masses, with the minimum water world mass being a function of the host star mass. This was attributed\cite{Burn2021arXiv210504596B} to the migration of icy planets from beyond the ice line into the inner regions of the disk. Inward migration becomes efficient at lower planetary masses around lower mass stars, which do not retain an envelope. For more massive stars, migration only occurs for planets above $10\,M_\oplus$ which are capable of accreting an envelope. Therefore, we propose that the observed population of planets around lower mass M dwarfs includes more ice-rich cores, with low masses and without envelopes. Our finding of a minimum mass for water worlds of $2\,M_\oplus$ is also in agreement with simulations \cite{Burn2021arXiv210504596B} for stellar host masses between $0.3$ and 0.5\,Solar masses ($M_\odot$) --- the majority of our sample. We conclude that rocky planets form within the ice line while water worlds (as defined in Fig.~1) formed beyond the ice line and migrated inwards. Our sample includes multi-planet systems with planets on either side the radius valley. For those systems, we find the innermost planets are always rocky and less massive, while the outermost belong to the water world population (Fig.~S17).

For solar-type stars, which are higher mass than M dwarfs, theoretical models predict similar results. Based on mass-radius relations, the planets larger than the radius valley have been identified\cite{Zeng2019PNAS..116.9723Z} as water worlds, and simulations using global planet formation and evolution models seem to support this hypothesis \cite{Venturini2020}. The simulations produce a bimodal distribution of core mass and composition, which agree with the observations. To explore whether our results can be extended from M dwarfs to solar-type stars, we attempted an analysis of known planets around F-, G-, and K-type stars. The results are shown in Fig.~S19. The planet distributions share some of the features of the STPM sample, however, the low number of precisely characterised small planets around these stellar types prevent us from drawing conclusions (see Supplementary Text).

We conclude that STPMs can be classified into three groups using their bulk densities.All three planet types could potentially be habitable if the right conditions are met \cite{Bolmont2017,Kite2018,Nikku2021}. However, determining those conditions from observations requires knowing the composition of these small planets.



\bibliography{fullbiblio}

\bibliographystyle{Science}

{\bf Acknowledgements}: 
We thank H.~Parviainen for help with the photometry analysis of the L~98-59 planetary system. 

{\bf Funding}: 
Supported by the Spanish Ministerio de Ciencia e Innovación, through project PID2019-109522GB-C52, the Centre of Excellence “Severo Ochoa”award SEV-2017-0709 to the Instituto de Astrofísica de Andalucía, and the University of La Laguna through the Margarita Salas Fellowship from the Spanish Ministry of Universities ref. UNI/551/2021-May-26 and under EU Next Generation funds (R.L.) and by the Spanish Ministry of Economics and Competitiveness through grant PGC2018-098153-B-C31 (E.P.). 

{\bf Author contributions}: 
R.L. analyzed and interpreted the data. E.P. interpreted the data. Both authors wrote the manuscript.

{\bf Competing Interests}: 
The authors declare that they have no competing interests.

{\bf Data Availability}: 
The Transiting M-dwarf Planets catalog (13) is available at \\ \textcolor{blue}{https://carmenes.caha.es/ext/tmp/}. Our STPM catalog, before and after our revision to the parameters of nine planets, is provided in data S1. The HARPS radial velocity observations we used are available in the ESO Science Archive Facility,\\ \textcolor{blue}{http://archive.eso.org/eso/eso\_archive\_main.html}, under target names GJ~1252 and LHS~1815. The TESS observations we used are available in the Mikulski Archive for Space Telescopes, \textcolor{blue}{https://archive.stsci.edu/tess/}, under target names L~98-59 and LTT~3780. Reduced transit photometry and radial velocity measurements for the systems revised in this work are also provided in data S1.
 
\section*{Supplementary materials}
Materials and Methods\\
Supplementary Text\\
Figures S1-S20\\
Tables S1-S8\\
Data S1\\
References \textit{(31-94)}

\clearpage

\begin{figure}
    \centering
    \includegraphics[width=0.49\hsize]{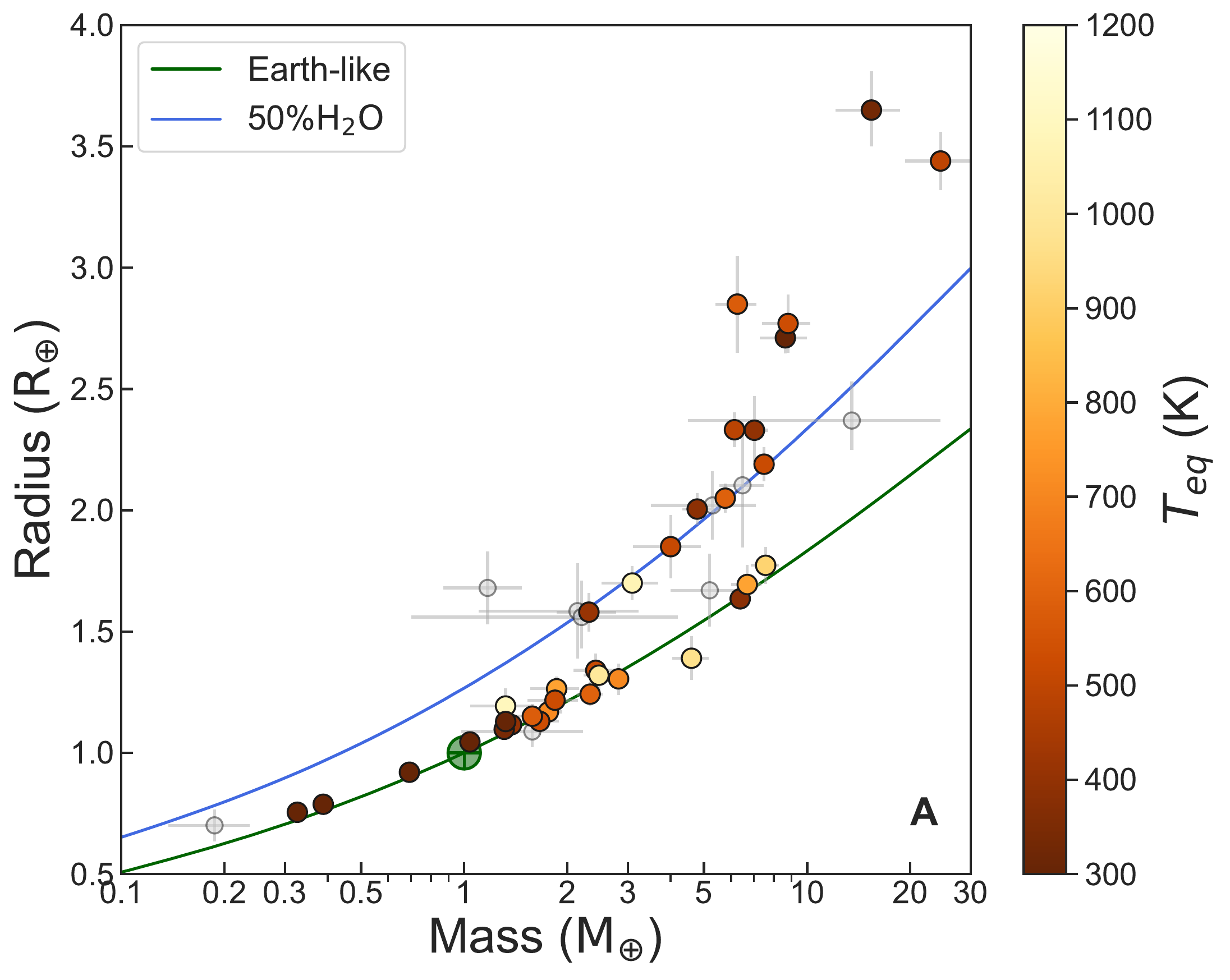}
    \includegraphics[width=0.49\hsize]{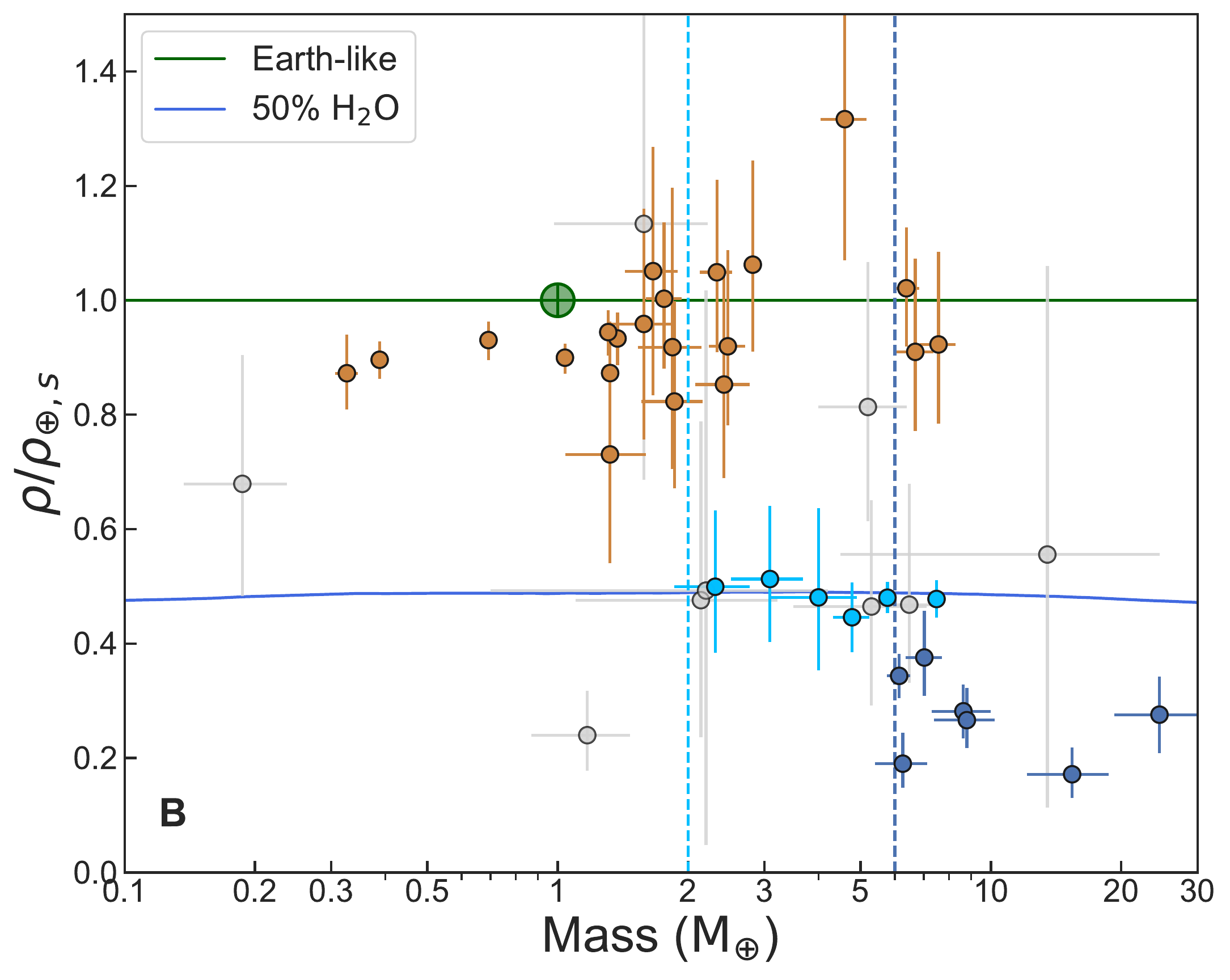}
    \caption{{\bf Mass-radius (A) and mass-density (B) diagrams for small transiting planets around M dwarfs (STPMs)}. Numerical values are provided in Data S1 and include nine planets with revised masses and radii (Table~S8). Error bars indicate 1$\sigma$ uncertainties of individual measurements. In both panels, two theoretical composition models\cite{Zeng2019PNAS..116.9723Z} are plotted: an Earth-like composition (32.5\% iron mass fraction and 67.5\% silicates, green curve) and a planet consisting of 50\% water-dominated ices and 50\% silicates (blue curve). In (\textbf{A}), planets are colour-coded by their equilibrium temperature $T_{\rm eq}$. In both panels, light grey points are planets with mass or radius determinations worse than our thresholds of 25\% and 8\%, respectively, so are not included in the subsequent analysis. In (\textbf{B}), densities are normalised by the Earth-like model and planets are colour-coded according to their characteristic bulk densities: rocky planets (brown), water worlds (light blue), and puffy sub-Neptunes (dark blue). The vertical dashed lines mark the $2 M_{\oplus}$ lower limit for water worlds (light blue) and $6 M_{\oplus}$ lower limit for puffy sub-Neptunes (dark blue). For reference, Earth is shown with a green $\oplus$ symbol}.
    \label{fig:mass-radius}
\end{figure}

\begin{figure}
    \centering
    \includegraphics[width=0.49\hsize]{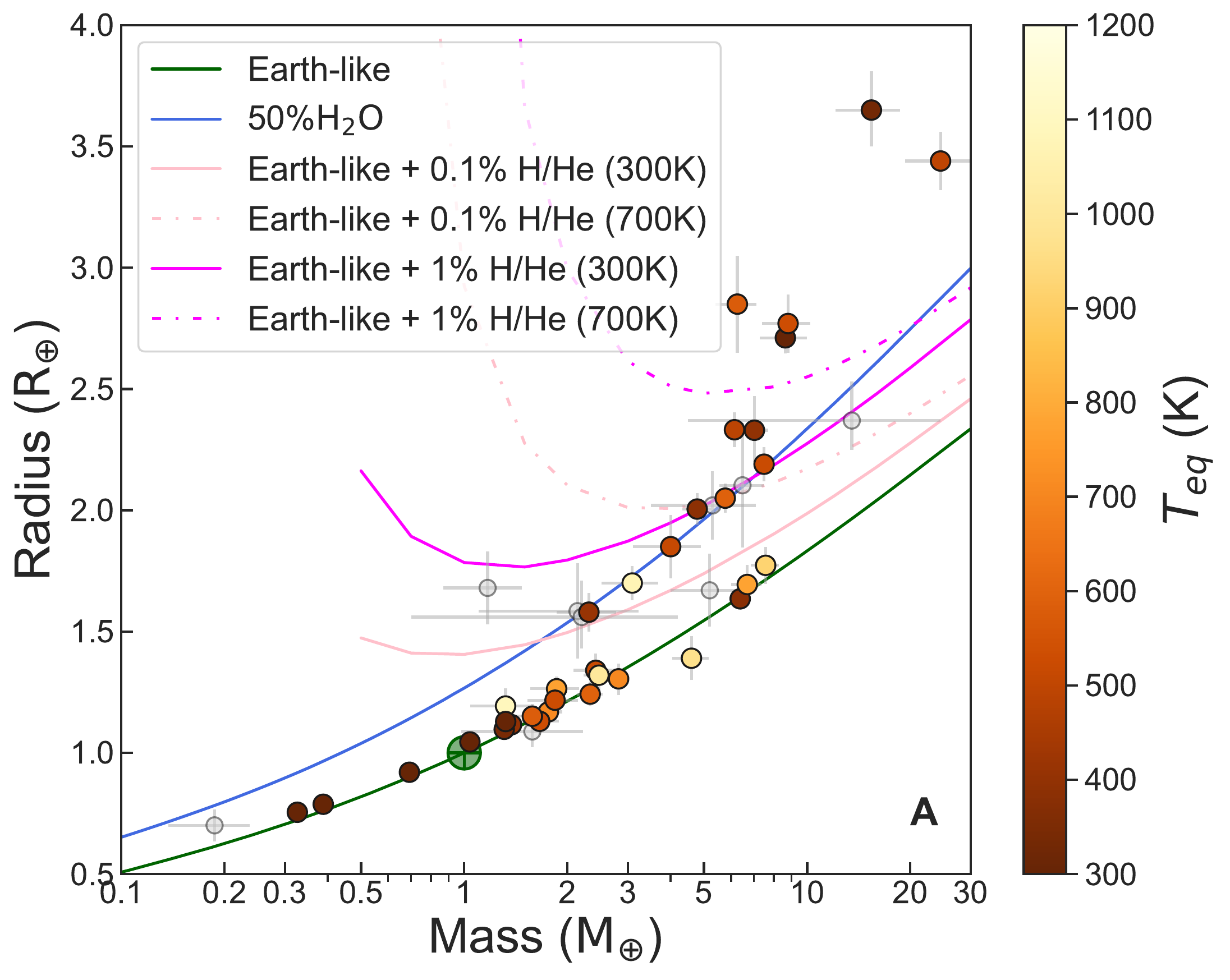}
    \includegraphics[width=0.49\hsize]{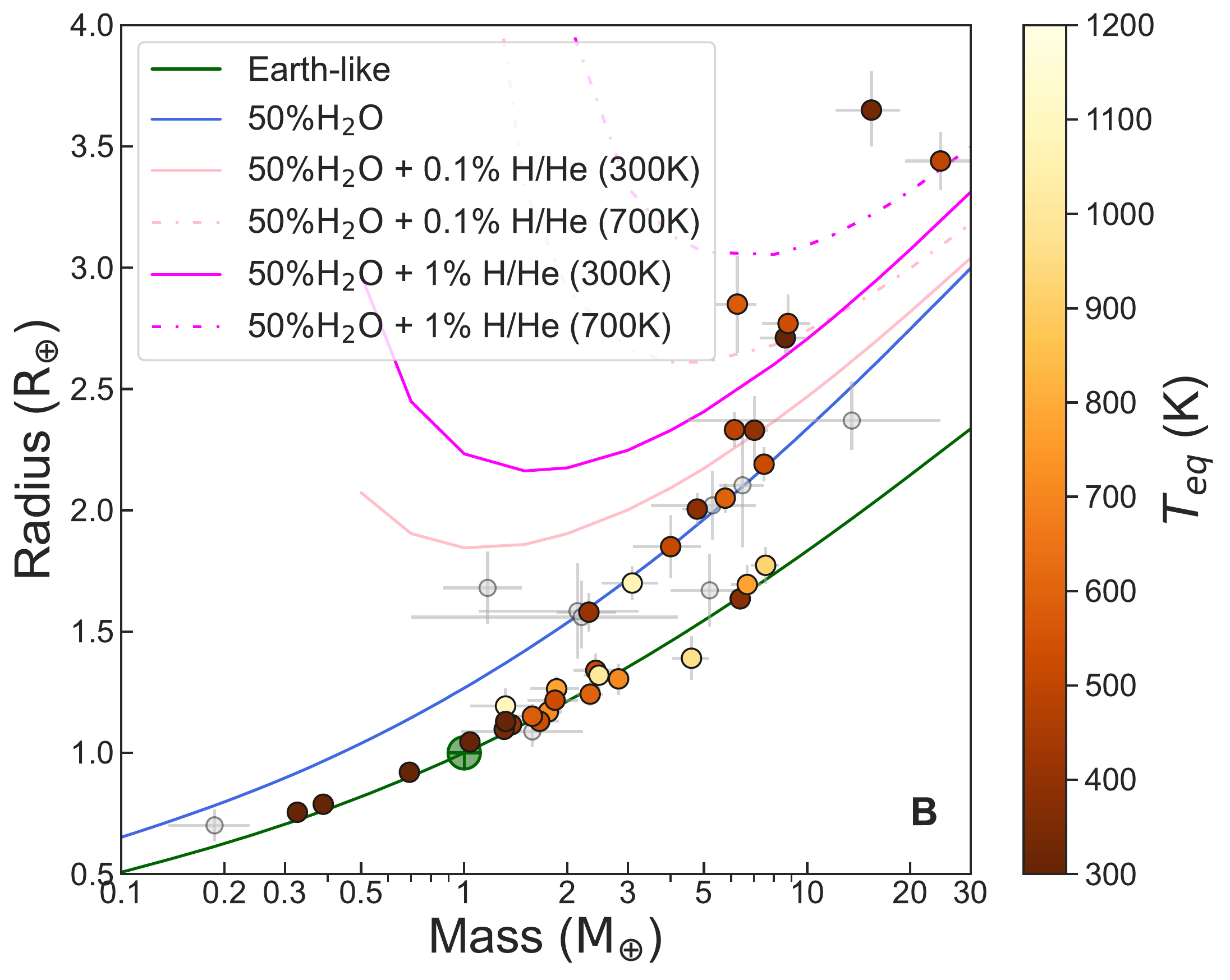}
    \includegraphics[width=0.49\hsize]{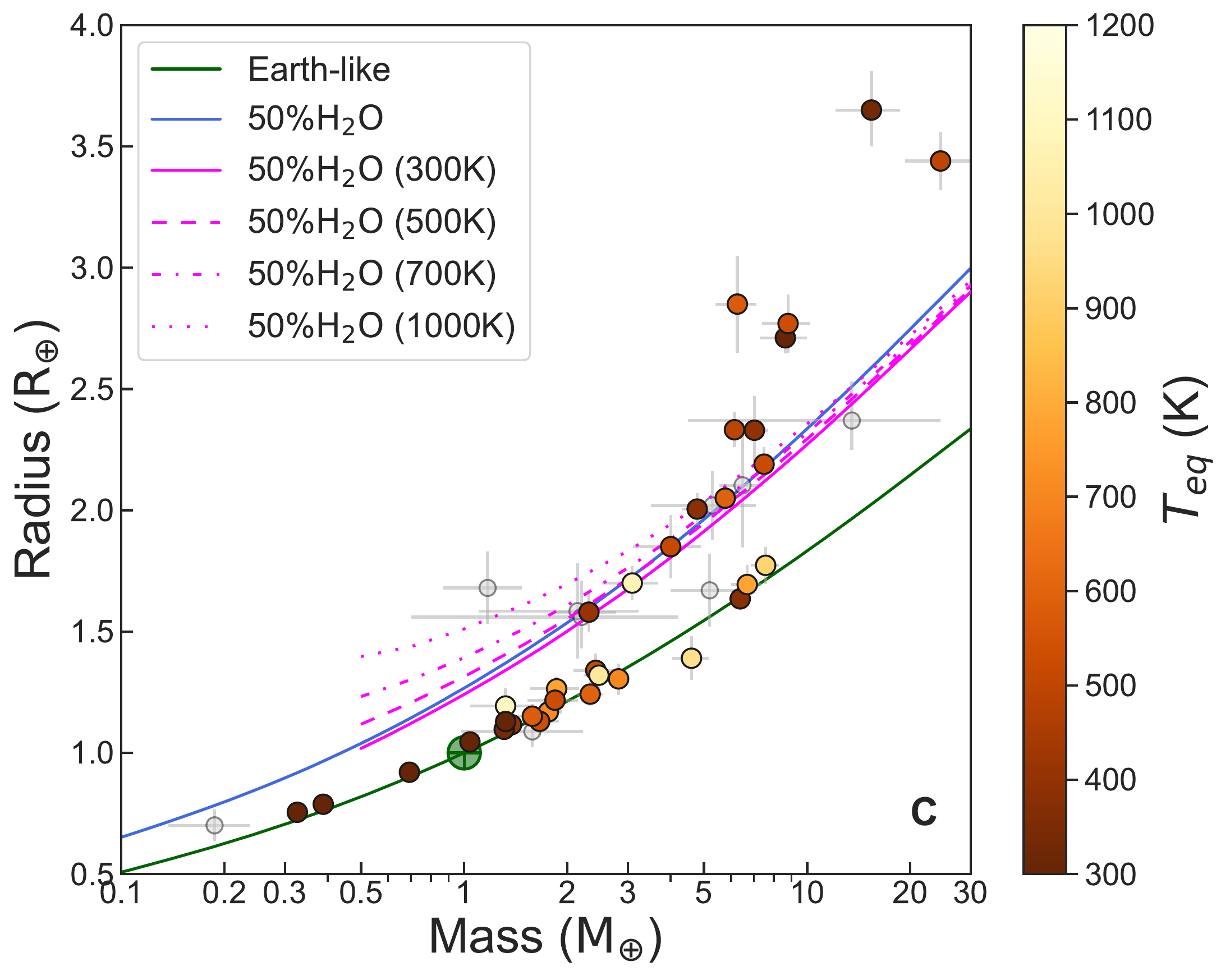}
    \includegraphics[width=0.49\hsize]{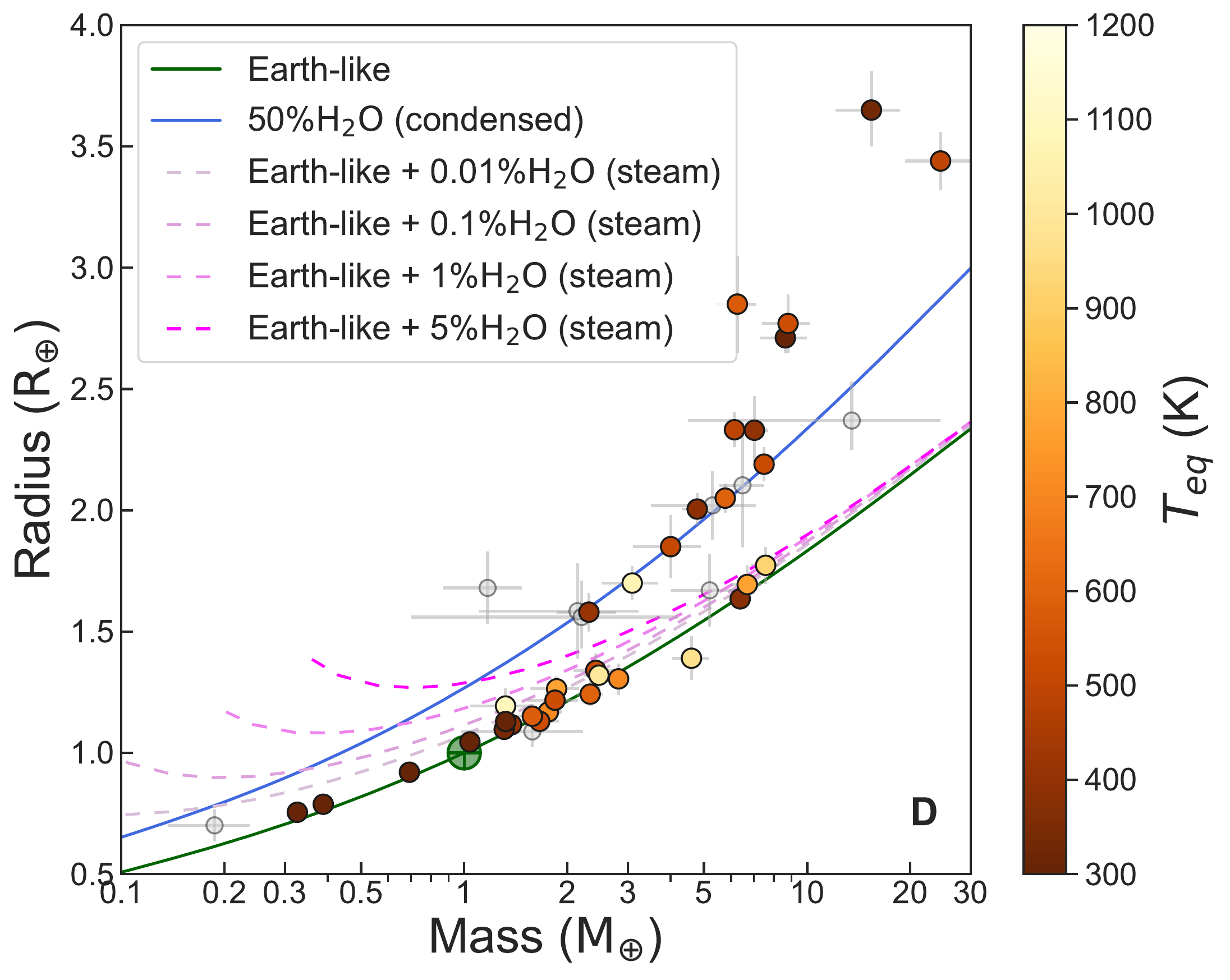}
    \caption{{\bf Same as Fig.~1A, but for different internal composition models}. (\textbf{A}): Earth-like rocky cores with H/He atmospheres by different percentages in mass at various temperatures \cite{Zeng2019PNAS..116.9723Z}. (\textbf{B}): Water-rich cores (50\% Earth-like rocky core plus 50\% water layer) with different mass fractions of H/He atmospheres at various temperatures \cite{Zeng2019PNAS..116.9723Z}. (\textbf{C}): Water worlds at different temperatures \cite{Zeng2019PNAS..116.9723Z}. (\textbf{D}): models for Earth-like planets accounting for runaway greenhouse radius inflation \cite{Turbet2020}.}
    \label{fig:mr-models}
\end{figure}

\begin{figure}
    \centering
    \includegraphics[width=\hsize]{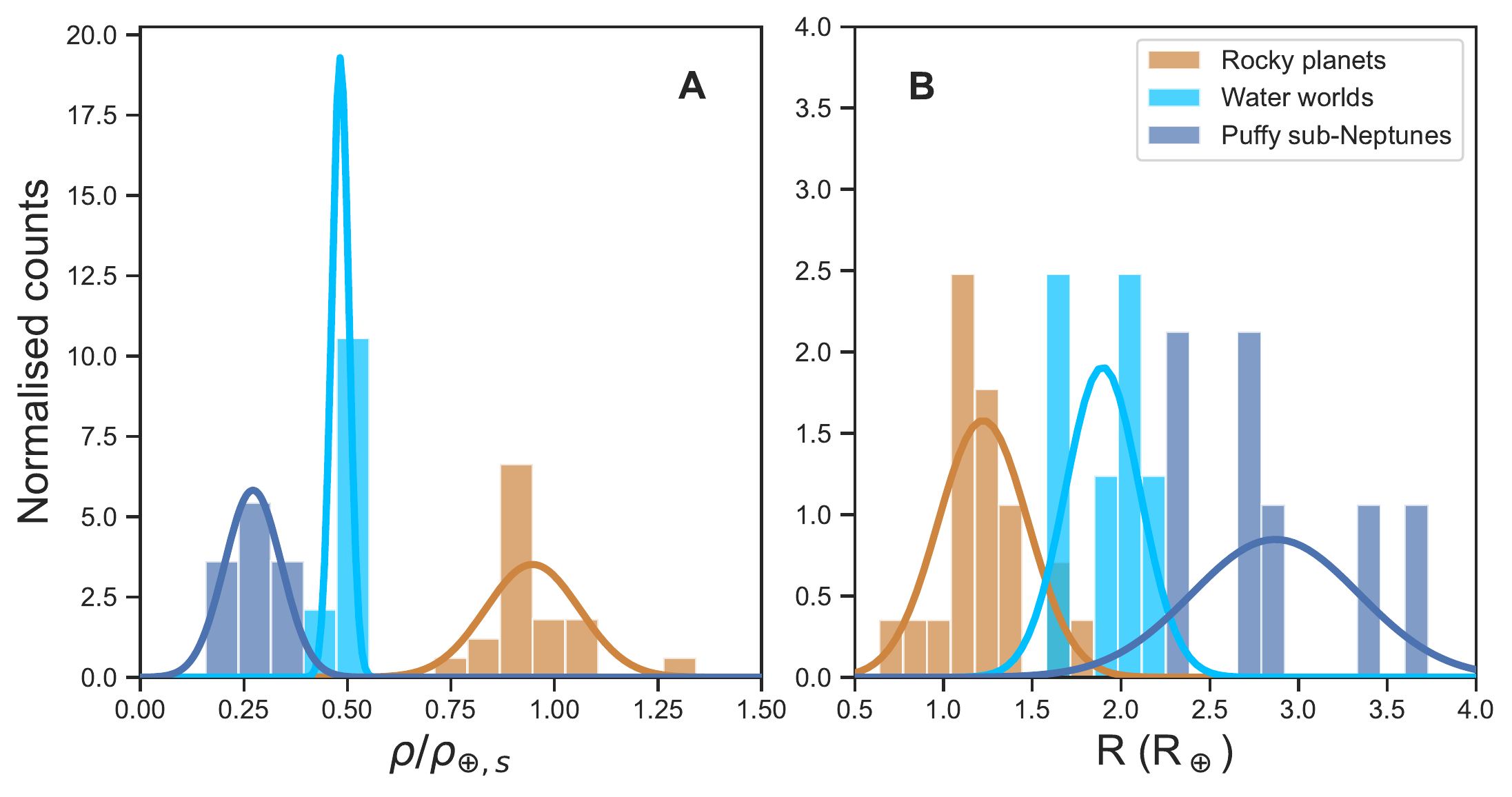}
    \caption{{\bf Normalised histograms of the STPM sample. Only planets that pass our precision requirements are included}. (\textbf{A}): Frequency as a function of density divided by an Earth-like model. (\textbf{B}): Frequency as a function of planetary radius. Colours as in Fig.~1B. Solid lines show Gaussian models fitted to the distribution of each planet type.}
    \label{fig:histogram}
\end{figure}

\begin{figure}
    \centering
    \includegraphics[width=0.67\hsize]{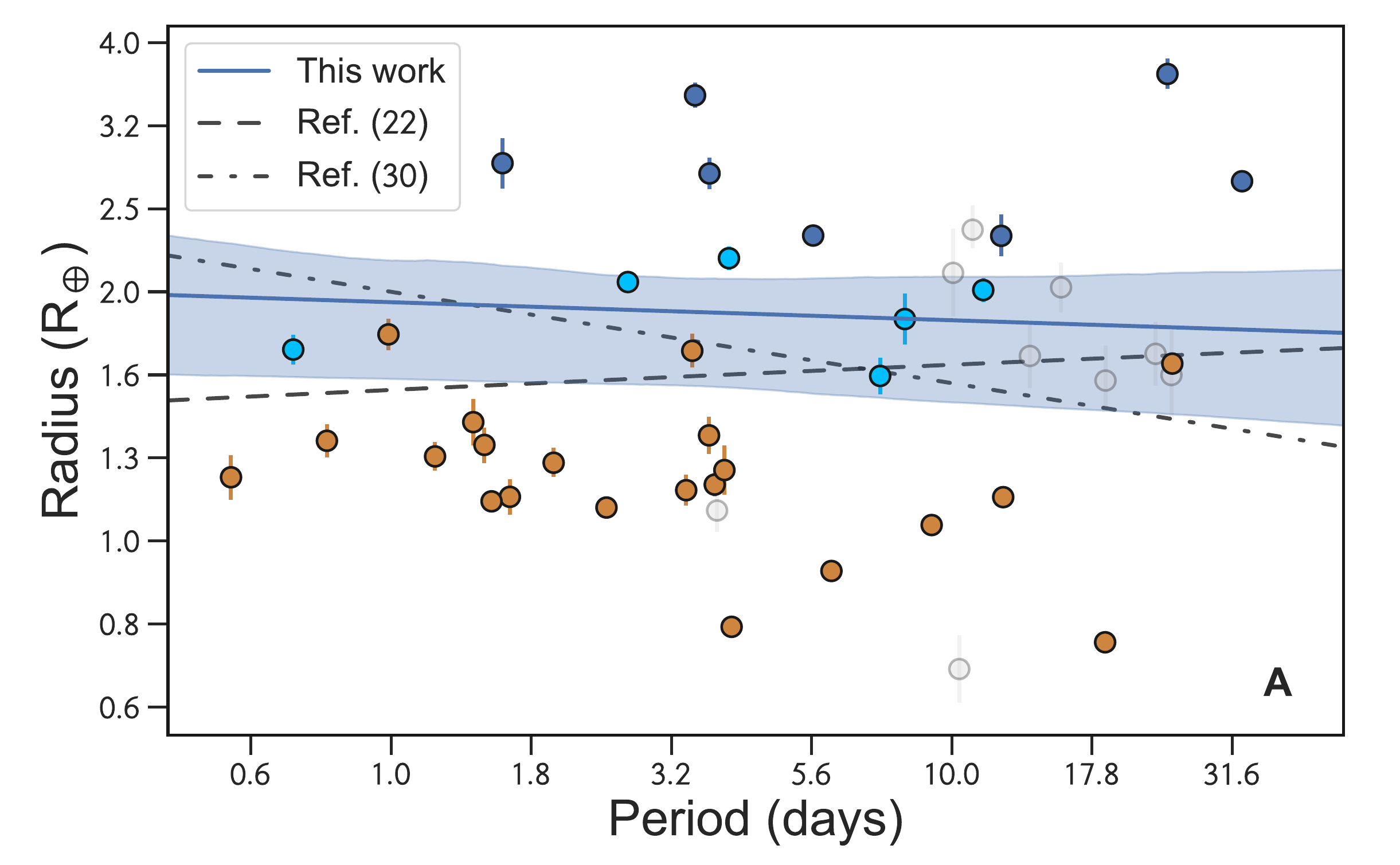}\\
    \includegraphics[width=0.67\hsize]{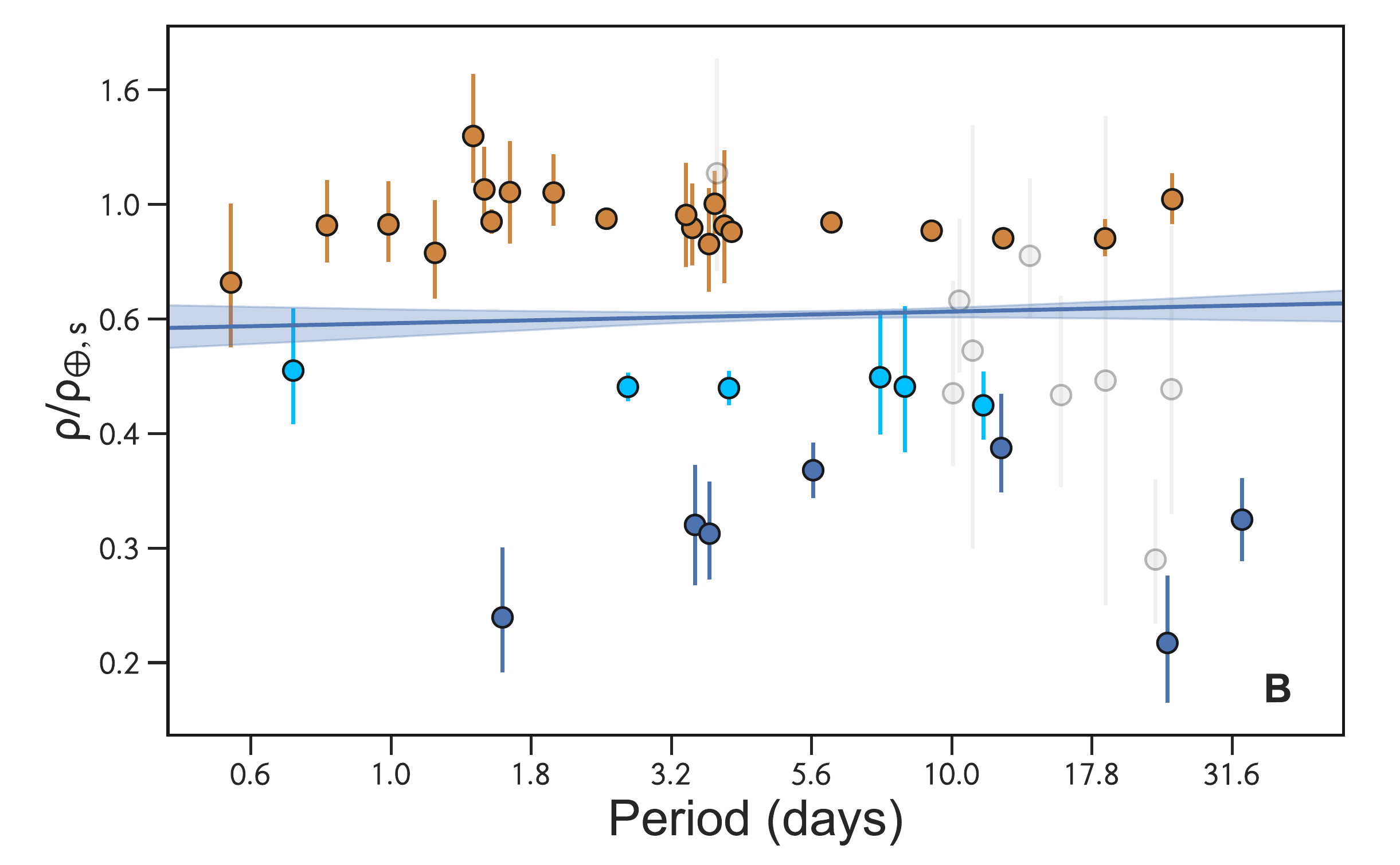}
    \caption{{\bf Radius (A) and density (B) as a function of orbital period for the complete STPM sample}.  Planets are colour-coded according to their bulk density as in Fig.~1B. The dashed lines are previous determinations of the location of the radius valley\cite{VanEylen2021,CloutierMenou2020AJ....159..211C}. The blue line and shaded region shows the best model and 1$\sigma$ uncertainties of the fit in this work, which is consistent with zero.}
    \label{fig:valley-insolation}
\end{figure}

\nocite{Trifonov2021Sci...371.1038T} 
\nocite{Southworth2011MNRAS.417.2166S}
\nocite{MartinezRodriguez2019ApJ...887..261M}
\nocite{TESS}
\nocite{Kostov2019}
\nocite{Nelson2020AJ....159...73N}
\nocite{2008ConPh..49...71T}
\nocite{CloutierMenou2020AJ....159..211C}
\nocite{Cloutier2021}
\nocite{Hirano2021}
\nocite{HARPS-N}
\nocite{IRD}
\nocite{Bluhm2021}
\nocite{Hirano2021}
\nocite{CARMENES2020}
\nocite{Gan2020}
\nocite{HARPS}
\nocite{SERVAL}
\nocite{LoCurto2015Msngr.162....9L}
\nocite{Kostov2019}
\nocite{Cloutier2019}
\nocite{Bluhm2020A&A...639A.132B}
\nocite{Cloutier2020}
\nocite{HIRES}
\nocite{Nowak2020}
\nocite{Cloutier2020a}
\nocite{iSHELL}
\nocite{Shporer2020}
\nocite{SERVAL}
\nocite{Otegi2020A&A...640A.135O}
\nocite{Zeng2019PNAS..116.9723Z}
\nocite{OwenWu2017ApJ...847...29O,JinMordasini2018ApJ...853..163J}
\nocite{Zeng2019PNAS..116.9723Z}
\nocite{Ikoma2000}
\nocite{Zeng2019PNAS..116.9723Z}
\nocite{Turbet2019,Turbet2020}
\nocite{Santos2017,Michel2020}
\nocite{Michel2020}
\nocite{Lopez2017}
\nocite{Dai2019}
\nocite{Raymond2018,Venturini2020}
\nocite{Venturini2020}
\nocite{Kurosaki2014,Lopez2017}
\nocite{CloutierMenou2020AJ....159..211C,VanEylen2021}
\nocite{Fulton17,VanEylen2018MNRAS.479.4786V,HardegreeUllman2020,Berger2020}
\nocite{VanEylen2018MNRAS.479.4786V,Martinez2019}
\nocite{LopezRice2018}
\nocite{VanEylen2021}
\nocite{VanEylen2018MNRAS.479.4786V,Martinez2019}
\nocite{OwenWu2017ApJ...847...29O,JinMordasini2018ApJ...853..163J}
\nocite{Ginzburg2018MNRAS.476..759G,GuptaSchlichting2019MNRAS.487...24G}
\nocite{VanEylen2021}
\nocite{gapfit}
\nocite{Southworth2011MNRAS.417.2166S}
\nocite{Cifuentes2020A&A...642A.115C}
\nocite{Luque2019A&A...628A..39L}
\nocite{Laughlin2004,Alibert2011}
\nocite{Zhu2018,Bryan2019}
\nocite{Henry2006}
\nocite{AlibertBenz2017A&A...598L...5A}
\nocite{Laughlin2004,IdaLin2005,Alibert2011}
\nocite{Raymond2007} 
\nocite{Schoonenberg2019,Coleman2019}
\nocite{Ormel2017,Liu2020,Brugger2020,Miguel2020}
\nocite{Burn2021arXiv210504596B}
\nocite{Venturini2020,Brugger2020,Izidoro2021}
\nocite{Venturini2020,Izidoro2021}
\nocite{Mayor2011arXiv1109.2497M,Mulders2018}
\nocite{WeissMarcy2014,Otegi2020}
\nocite{Lissauer2011ApJS..197....8L,Fabrycky2014ApJ...790..146F}
\nocite{Bitsch2021,Schlecker2020arXiv200705563S}
\nocite{Morales2019,Sabotta2021}
\nocite{Bitsch2021}
\nocite{Venturini2020}

\end{document}